\documentclass[doublecol]{epl2} 

\title{Roles of mutation rate and co-existence of multiple strategy updating rules in evolutionary prisoner's dilemma games}
\shorttitle{Mutation and multiple strategy updating rules in social dilemma} 

\author{Hirofumi Takesue\inst{1}}
\shortauthor{H. Takese}

\institute{Faculty of Political Science and Economics, Waseda University - 1--104, Tozuka, Sinjuku, Tokyo, 1698050, Japan}


\abstract{
The emergence and maintenance of cooperation has attracted intensive scholarly interest and has been analysed within the framework of evolutionary game theory. The role of innovation, which introduces novel strategies into the population, is a relatively understudied aspect of evolutionary game theory. Here, we investigate the effects of two sources of innovation---mutation and heterogeneous updating rules. These mechanisms allow agents to adopt strategies that do not rely on the imitation of other individuals. The model introduces---in addition to canonical imitation-based strategy updating---aspiration-based updating, whereby agents switch their strategy by referring solely to the performance of their own strategy; mutation also introduces novel strategies into the population. Our simulation results show that the introduction of aspiration-based rules into a population of imitators leads to the deterioration of cooperation. In addition, mutation, in combination with heterogeneous updating rules, also diminishes cooperators. This phenomenon is prominent when a large proportion of the population consists of imitators rather than adopters of aspiration-based updating. Nevertheless, a high mutation rate, in combination with a low aspiration level, has positive nonlinear effects, and a heterogeneous population achieves a higher level of cooperation than the weighted average of homogeneous populations. Our results demonstrate the profound role of innovation in the evolution of cooperation.}

\begin{document}

\maketitle

\section{Introduction}
A core problem in societies is the tension between individual selfishness and social efficiency. While mutual cooperation leads to the socially efficient outcomes, individuals are tempted to take advantage of others' cooperative efforts without bearing the cost of cooperation \cite{Hardin1968}. To understand the manner in which cooperation is maintained and the reason for its wide prevalence in societies despite this susceptibility to `free-riding', intensive research, adopting the framework of evolutionary game theory, has been widely conducted in the biological and social sciences \cite{Axelrod1984, Nowak2006}. Statistical physicists have also extracted in-depth understanding of this topic through the application of their analytical and computational methods \cite{Szabo2007a, Roca2009, Perc2017, Sanchez2018}. 

Network structure is a cooperation-enhancing mechanism that has attracted much attention. Here, instead of interacting with randomly selected individuals, agents in the system interact with (fixed) neighbours in networks. Since Nowak and May's \cite{Nowak1992} seminal work showing that a lattice structure supports cooperation, several studies have examined the relationship between cooperation and many types of complex networks \cite{Santos2005, Szolnoki2008, Roca2009a}. Many investigations have interrogated the role of network structure in combination with other mechanisms, including network evolution \cite{Zimmermann2004, Fu2007, Szolnoki2008b, Perc2010, Wardil2014}; punishment \cite{Helbing2010c, Szolnoki2013d, Chen2014, Chen2015c, Chen2015, Perc2015, Yang2015b, Ohdaira2016, Ohdaira2017, Takesue2018}; mobility \cite{Yang2010, Lin2011}; the mixing of different games \cite{Amaral2015, Amaral2016a}; exclusion from social games \cite{Li2015a, Li2016d, Szolnoki2017}; strategy persistence \cite{Huang2017b, Huang2018f}; overconfidence \cite{Szolnoki2018b} and environmental feedback \cite{Szolnoki2017b, Chen2018c}.

Payoff-based copying of strategies is the basic component of evolutionary games. Many studies assume that agents compare their payoff (fitness) with that of their neighbour, copying that neighbour's strategy with higher probability if that neighbour achieves a higher payoff \cite{Szabo1998}. This rule may be understood as the model of genetic evolution or cultural evolution through social learning (imitation) \cite{Rand2013}. This mechanism depends on other individuals as the copy model, and once a strategy becomes extinct, it will not reemerge in the population.

Here, we aim to examine the role of \textit{innovation} in evolutionary processes, where innovation connotes the adoption of a strategy that does not rely on others as an information source. In particular, we investigate the combined effects of two sources of innovation: heterogeneous updating rules and mutation. We first address the updating rules that determine the manner in which agents modify their strategies. In addition to the widely adopted payoff-based copying of a neighbour's strategy \cite{Szabo2007a}, many studies have examined the outcomes of other updating rules such as aspiration dynamics \cite{Chen2008, Zhang2011a, Yang2012a, Du2014, Amaral2016}; Glauber dynamics \cite{Roca2009b, Li2017b, Amaral2017, Amaral2018, Danku2018} and conformity \cite{Pena2009, Cui2013, Szolnoki2015, Javarone2016a, Shu2019}. 

Whether a rule is innovative or non-innovative is one important classification criterion of updating rules \cite{Amaral2018}. Agents who use innovative dynamics can adopt a strategy that does not exist in the current population; in contrast, agents who use a non-innovative rule can only adopt strategies that exist and have been adopted by other agents in the current population. Because agents cannot copy specific strategies that have not already been used by any other agents, the widely adopted payoff-based copying of strategy is a non-innovative rule. The aspiration dynamics that we adopt in this study are an example of innovative dynamics. Agents who rely on this rule tend to maintain the current strategy if the strategy brings them a larger payoff than their aspiration level; in contrast, agents who are dissatisfied with the performance of their current strategy may try other strategies without referring to the strategies of others.

The multiple candidates for strategy-updating rules motivate researchers to investigate the effect of heterogeneous rules. Research has shown that heterogeneous dynamics lead to different cooperation levels and complex patterns that are not observed in homogeneous systems. For instance, the co-existence of two different payoff-based copying rules may enhance cooperation \cite{Szolnoki2018c}. In a well-mixed population, combination with an innovative rule can support cooperation among imitators \cite{Liu2016b}, whereas in structured population, it can deter cooperation \cite{Xu2017a, Amaral2018}. Here, we consider a model in which imitation dynamics and aspiration dynamics co-exist in the population.

Another source of innovation is mutation, in which individuals adopt one strategy randomly. Random modifications of traits may sometimes greatly alter the results of socio-physical models \cite{Macy2015}. Studies of evolutionary games have shown the vulnerability of cooperation sustained by the payoff-based copying rule to the introduction of mutation \cite{Helbing2009, Ichinose2018a}. Despite the importance of the effect of mutation in evolutionary games \cite{Adami2016}, this effect is understudied, particularly when the model includes innovative updating rules. A recent study has demonstrated the robustness of outcomes of evolutionary games against the network structure when agents adopt aspiration dynamics \cite{Du2015}. Robustness against mutation may also be an important factor in aiding our understanding of the applicability of updating rules.

Here, we examine the effects of the mutation rate and heterogeneous updating rules, and investigate the role of innovation in the prisoner's dilemma games. In the second and third sections, we explain the scrutinised model and explain the results of the Monte Carlo simulations, respectively. In the final section, we discuss the implication of our model and its possible future extension. 

\section{Model}
We consider the social games played by agents located on the $L \times L$ square lattice with a periodic boundary condition. All agents have eight neighbours (Moore neighbourhood), with whom they participate in games. In examining the role of the extension of simple models, the adoption of the lattice structure is useful \cite{Perc2017}. Heterogeneous networks, unlike regular networks, can affect evolutionary outcomes through not only the different numbers of games that agents participate in \cite{Santos2005} but also the different numbers of opportunities to be referred as a role agent in the imitation process \cite{Huang2017, Takesue2019}.

We adopted prisoner's dilemma games as a simple model of the social dilemma. Agents have two strategies in this game: cooperation and defection. Mutual cooperation achieves payoff $R=1$, whereas mutual defection leads to payoff $P=0$. If one agent chooses cooperation (defection) and his/her partner chooses defection (cooperation), she/he will gain $S=0$ ($T=b$). Here, the harshness of the social dilemma is controlled by one parameter---$b$ ($1 < b < 2$) \cite{Nowak1992}. Initially, half of the agents adopt cooperation.

Each round involves agents participating in the game with all of their neighbours and accumulating payoff. Agents update their strategy on the basis of their payoff. In this study, following the seminal article involving aspiration dynamics, we assume synchronous updating \cite{Chen2008}. Imitation dynamics and aspiration dynamics are the two types of dynamics co-existing in the model. Throughout a simulation run, each agent keeps adopting one of these two rules. At the outset of simulation, $p L^2$ agents are assigned the aspiration-based rule, and the rest are assigned the imitation rule. 

In strategy updating by imitation dynamics, the focal agent ($i$) chooses one of the neighbours randomly as a role agent ($j$). The accumulated payoffs of these two agents ($\Pi_i$ and $\Pi_j$) are compared, and the focal agent tends to imitate the role agent's strategy when the role agent earns a larger payoff. The imitation probability is determined by the following equation \cite{Szabo1998}:
\begin{equation}
\mathrm{P}_{i \leftarrow j} = 1/[1 + \exp(\beta(\Pi_i - \Pi_j))],
\end{equation}
where $\beta$ is the intensity of the effect of payoff difference on the imitation probability ($\beta \rightarrow \infty$ implies deterministic imitation, whereas $\beta \rightarrow 0$ implies random adoption of the strategy).

Other agents update their strategy according to the aspiration-based rule. An agent who adopts the aspiration-based rule ($i$) compares his/her payoff with an exogenously set aspiration level ($A$). The focal agent is more likely to adopt another strategy when the current strategy does not realise a satisfying payoff. The probability of switching is calculated by the following equation \cite{Chen2008}:
\begin{equation}
\mathrm{P}_{iA} = 1/[1 + \exp(\beta(\Pi_i - k_i A))],
\end{equation}
where $k_i$ (agents' degree) is set at 8 in this study. For the sake of simplicity, we utilised the same values of $\beta$ for imitation dynamics and aspiration dynamics. 

The agents' strategy can be modified by not only the payoff-based strategy updating but also mutation. Mutation occurs with the probability of $\mu$, and when it does, agents adopt one strategy randomly. Our interest is in understanding the manner in which the mutation rate ($\mu$) affects the evolutionary outcomes. 

\section{Results}
In our investigation of evolutionary outcomes, we conducted Monte Carlo simulation. Simulation runs first continued for $10^4$ periods, and we subsequently sampled the average of the following $2000$ periods. We conducted four simulation runs for each combination of parameters and reported their average values, focussing our principal interest on the proportion of cooperators in stationary states ($\rho_C$). We also observed $\rho_C^I$ ($\rho_C^A$), the proportion of cooperators among the agents who adopted imitation (aspiration) dynamics. In addition, we sometimes investigated the presence of nonlinear effects of heterogeneity in strategy updating rules. In this case, we examined the following quantity \cite{Amaral2018}:
\begin{equation}
\Delta \rho = \rho_{\rm{mix}} - [(1-p)\rho_{\rm{imt}} + p\rho_{\rm{asp}}],
\end{equation}
where $\rho_{\rm{mix}}$ is the proportion of cooperators in the heterogeneous system in which $(1-p) L^2$ agents adopt imitation dynamics and $pL^2$ agents adopt aspiration dynamics. In contrast, $\rho_{\rm{imt}}$ ($\rho_{\rm{asp}}$) is the outcome with a homogeneous population where all agents adopt imitation (aspiration) dynamics. If the two rules do not have mutual influence, the resultant cooperation level will be merely the weighted average of two homogeneous populations (i.e., $\Delta \rho$ is 0) because compared cases share the same parameter values. Hence, we can adopt this quantity as another indicator of the success of cooperators \textit{given the specific value of $p$}.

\begin{figure}[tb]
\centering
\vspace{0mm}
\includegraphics[width = 85mm, trim= 0 10 0 0]{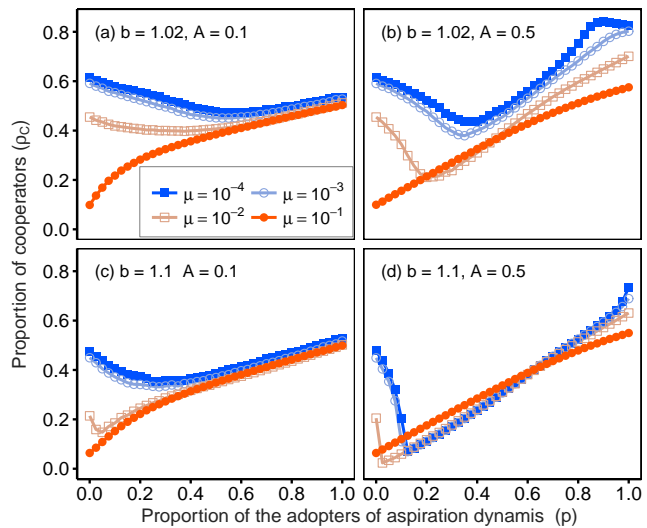}
\caption{\small The proportion of cooperators as a function of the proportion of adopters of aspiration dynamics ($p$). Drops in cooperation were observed with intermediate values of $p$ as long as $\mu$ was not overly large. Mutation essentially diminished the proportion of cooperators, and populations with large proportions of imitators were vulnerable to mutation. A high mutation rate ($\mu = 0.1$) and a low aspiration level ($A = 0.1$) enhanced cooperation relative to the weighted average of homogeneous systems. Fixed parameters: $L = 100, \beta = 10$.}
\label{p_mu_b_A}
\end{figure}

First, we present an overview of the basic patterns of the simulation results. Figure \ref{p_mu_b_A} reports the proportion of cooperators as a function of $p$. In many cases, the introduction of aspiration dynamics into a homogeneous population of imitators negatively affected the cooperation level. This pattern was also observed in previous literature examining imitation and innovative dynamics in a heterogeneous system. These studies attributed this trend to the destruction, by the introduction of \textit{innovative} dynamics, of the cluster of cooperators that could have fostered cooperators' expansion \cite{Xu2017a, Amaral2018}. As a result of this drop, in many cases, as demonstrated in the figure, the cooperation level was lower than the weighted average of homogeneous systems (i.e., $\Delta\rho$ is negative).

In addition, overall patterns of the figure show that a high mutation rate essentially leads to lower cooperation levels. Populations with a large proportion of imitative agents were particularly vulnerable to mutation. In contrast, a relatively high cooperation level was maintained when a large proportion of the population adopted aspiration dynamics. At the same time, a combination of frequent mutation and heterogeneous dynamics could have positive nonlinear effects on cooperation. We observed clear concave relationships between $p$ and $\rho_C$ with a high mutation rate ($\mu = 0.1$) and a low aspiration level ($A = 0.1$), which implies that $\Delta \rho$ was positive with these parameter values.

\begin{figure*}[tb]
\centering
\vspace{0mm}
\includegraphics[width = 125mm, trim= 0 10 0 0]{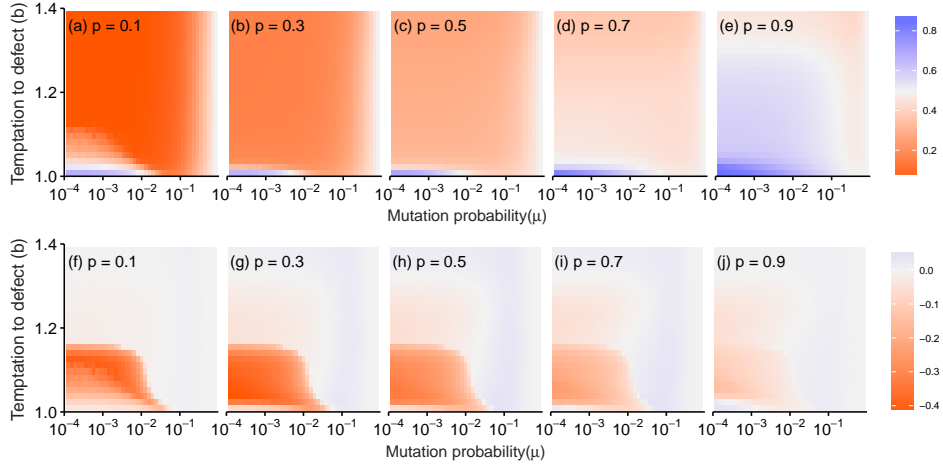}
\caption{\small The proportion of cooperators ($\rho_C$; panel (a)--(e)) and the effects of heterogeneity ($\Delta \rho$; panel (f)--(j)) as a function of the mutation probability ($\mu$) and temptation to defect ($b$). As long as the values of $\mu$ were not overly large, mutation diminished cooperation levels. A population with a large number of adopters of imitation (aspiration) dynamics is vulnerable (robust) to (against) mutation. Although we ran the simulation with $b \leq 1.8$, we report the results for $b \leq 1.4$ because the results were stable with larger $b$. Fixed parameters: $L = 100, \beta = 10, A = 0.5$.}
\label{phase_mu_b_A05}
\end{figure*}

Expanding on the simulation results, the upper (lower) panel of fig. \ref{phase_mu_b_A05} reports $\rho_C$ ($\Delta \rho$) as a function of $\mu$ and $b$ for different values of $p$. In this figure, the value of $A$ is set at 0.5. First, the figure clearly shows that the prevalence of cooperators was broken by the introduction of mutation. Although we observed a large number of cooperators with a small $b$, the proportion of cooperators diminished as the value of $\mu$ grew. This tendency was reversed only after $\mu$ became overly large, and the system approached totally random results. In addition, the parameter regions in which cooperators could achieve a high frequency depended on the value of $p$. We observed a similar pattern with $\Delta \rho$. For instance, although the results with $p = 0.1$ (panel (f)) show small nonlinear effects in a parameter region of the bottom-left corner, this region is reduced when $p = 0.3$ (panel (g)). 

Second, a population occupied by a large number of adopters of aspiration dynamics is robust against mutation. Although the introduction of aspiration dynamics lowered the cooperation level (panels (a) and (b)) and had detrimental nonlinear effects (panels (f) and (g)), further increases in the numbers of agents adopting aspiration dynamics led to a higher cooperation level (panels (c)--(e)) and smaller detrimental effects of heterogeneous dynamics (panels (h)--(j)). When the number of defectors increased, more adopters of aspiration dynamics shifted to cooperation because they became unsatisfied with the low payoffs achieved by mutual defection. Consequently, a total breakdown of cooperation among agents using the aspiration-based rule was prevented. In contrast, imitation dynamics lacked such resilience and cooperators just decreased the frequency once their clusters were broken down by mutation.

\begin{figure}[tb]
\centering
\vspace{0mm}
\includegraphics[width = 87mm, trim= 0 10 0 0]{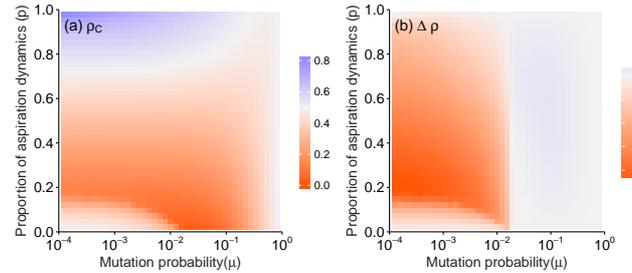}
\caption{\small The proportion of cooperators ($\rho_C$; panel (a)) and effects of heterogeneity ($\Delta \rho$; panel (b)) as a function of the mutation rate ($\mu$) and the proportion of the adopters of aspiration dynamics ($p$). Collaboration between mutation and heterogeneous dynamics lowered $\rho_C$. This tendency was prominent among populations with a large number of imitators. $\Delta \rho$ became almost zero once the values of $\mu$ became large enough. Fixed parameters: $L = 100, \beta = 10, b = 1.05, A = 0.5$.}
\label{phase_mu_p}
\end{figure}

We then varied the value of other parameters to scrutinise the observed patterns. Panel (a) of fig. \ref{phase_mu_p} reports $\rho_C$ as the function of $\mu$ and $p$. The figure demonstrates the hindering of cooperation by the collaboration of heterogeneous rules and a high mutation rate. When $p$ was less than approximately 0.2, we found that the level of $p$ required to suppress cooperation became smaller as $\mu$ became larger, suggesting that heterogeneous dynamics and mutation collaboratively caused cooperation to deteriorate. As noted above, innovative dynamics can hinder the formation of clusters of cooperators. Mutation, by introducing defectors, can also break these clusters. Consequently, the required number of adopters of aspiration dynamics drops as $\mu$ increases. A similar collaborative impediment of cooperation was observed with a large $p$ as well. Again, cooperation was inhibited to a smaller extent than it would in a population with large numbers of imitators. Panel (b) shows the values of $\Delta \rho$ in the same manner. The clearly observed pattern is that the values of $\Delta \rho$ are almost zero when mutation occurs frequently. Here, the introduction of aspiration dynamics no longer causes decrease in $\rho_C$ which corresponds to negative values of $\Delta \rho$.

\begin{figure}[tb]
\centering
\vspace{0mm}
\includegraphics[width = 85mm, trim= 0 10 0 0]{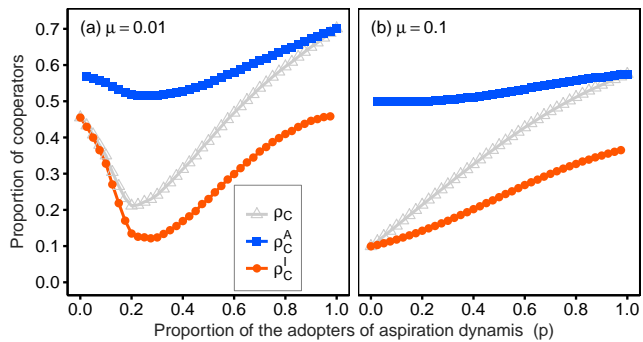}
\caption{\small The proportion of cooperators as a function of the proportion of agents adopting aspiration dynamics ($p$). Heterogeneous dynamics had larger effects on imitators. A higher frequency of mutation alone caused cooperation among imitators to deteriorate, and the relationship between $p$ and $\rho_C$ became linear. Fixed parameters: $L = 100, \beta = 10, b = 1.02, A = 0.5$.}
\label{p_mu_b_A_grp_A05}
\end{figure}

Observing the proportion of cooperators among agents who adopt each updating rule clarifies the roles of heterogeneous dynamics and the mutation rate. Figure \ref{p_mu_b_A_grp_A05} reports $\rho_C$, $\rho_C^A$ and $\rho_C^I$ as a function of $p$. Panel (a) shows that a mixture of heterogeneous dynamics had a larger influence on imitators ($\rho_C^I$) although heterogeneity also lowered $\rho_C^A$. Consequently, particularly in populations with a large number of imitators, we observed a drop in $\rho_C$ (and $\Delta \rho$) as shown in fig. \ref{phase_mu_b_A05}. When a higher mutation rate was introduced to the population ($\mu = 0.1$), however, fast mutation alone, without the introduction of aspiration dynamics, could cause a deterioration in cooperation among imitators. Here, the relationship between $p$ and $\rho_C$ approached a linear one, and $\Delta \rho$ approached zero as we observed in fig. \ref{phase_mu_p}(b). 

\begin{figure*}[tb]
\centering
\vspace{0mm}
\includegraphics[width = 125mm, trim= 0 10 0 0]{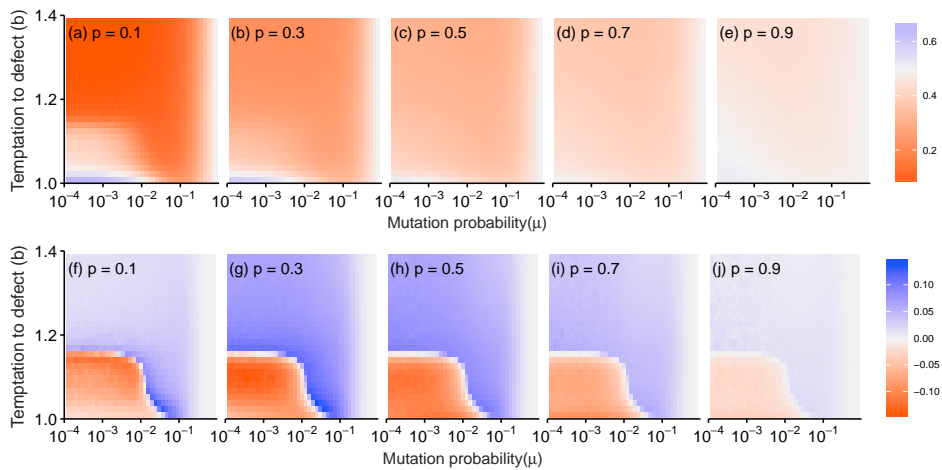}
\caption{\small The proportion of cooperators ($\rho_C$; panels (a)--(e)) and the effect of heterogeneity ($\Delta \rho$; panel (f)--(j)) as a function of the mutation probability ($\mu$) and temptation to defect ($b$). A combination of mutation and heterogeneous updating rules could have nonlinear positive effects on cooperation levels. Although we ran the simulation with $b \leq 1.8$, we reports the results for $b \leq 1.4$ because the results were stable with larger $b$. Fixed parameters: $L = 100, \beta = 10, A = 0.1$.}
\label{phase_mu_b_A01}
\end{figure*}

The results of the simulation thus far showed mainly that (the combination of) heterogeneous dynamics and mutation diminished the cooperation level. We then examined the system with other values of parameters to investigate whether we would observe a different pattern. Figure \ref{phase_mu_b_A01} reports the same results as fig. \ref{phase_mu_b_A05}, with a different aspiration level ($A = 0.1$). Although overall patterns are similar to those of the results in fig. \ref{phase_mu_b_A05} regarding the cooperation level ($\rho_C$), the figure also shows a different pattern: heterogeneous system often showed large nonlinear effects (i.e. positive $\Delta \rho$) and achieved higher cooperation level than the weighted average of homogeneous populations. The bottom panels of fig. \ref{phase_mu_b_A01} show that the values of $\Delta \rho$ were often larger than 0.1. This result was not observed in fig. \ref{phase_mu_b_A05}, suggesting that differences in aspiration levels led to differences in behaviour.

The positive $\Delta \rho$ with lower aspiration levels could be accounted for by the fact that adopters of aspiration dynamics will stably choose the same strategy despite the small payoff. Low aspiration induces agents to be satisfied with a small payoff, providing them with little motivation to change their behaviour. Introduced adopters of aspiration dynamics can remain cooperative even when the mutation frequency is high enough (or the temptation to defect is large enough) to cause total deterioration of the cooperation level among imitators. In addition, adopters of aspiration dynamics can function as \textit{stably} cooperative neighbours enhancing cooperation among imitators. Although a high aspiration level induces agents to choose cooperation with probability about 0.5, the frequent strategy switching prevents the aspiration dynamics from supporting cooperative imitators. 

\begin{figure}[tb]
\centering
\vspace{0mm}
\includegraphics[width = 85mm, trim= 0 10 0 0]{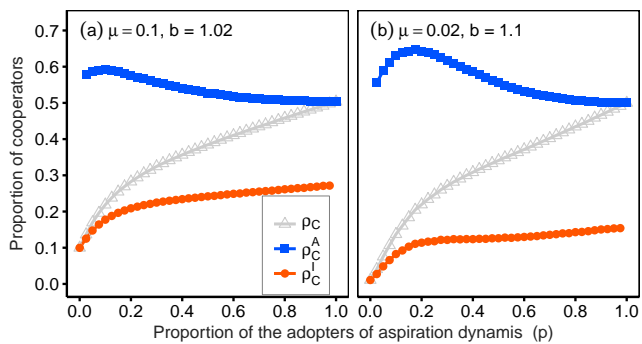}
\caption{\small The proportion of cooperators as a function of the proportion of adopters of aspiration dynamics ($p$). Agents who adopted aspiration dynamics chose cooperation even when the majority of imitators chose defection, thereby also enhancing cooperation among imitators. Fixed parameters: $L = 100, \beta = 10, A = 0.1$.}
\label{p_mu_b_A_grp_A01}
\end{figure}

Figure \ref{p_mu_b_A_grp_A01} reports $\rho_C$, $\rho_C^A$ and $\rho_C^I$ as a function of $p$ to corroborate this intuition. The values of $b$ and $\mu$ in these figures were selected so that a homogeneous population of imitators could achieve only low cooperation level (see $\rho_C^I$ when $p = 0$). In contrast to these low values of $\rho_C^I$, among the adopters of aspiration dynamics introduced into the population, cooperation was robustly maintained. Here, the introduction of agents who adopted the aspiration-based rule also supported cooperation among imitators. We observed a significant increase in $\rho_C^I$, suggesting that imitators also benefit from the introduction of aspiration dynamics. The harmful effect of the introduction of aspiration dynamics almost disappeared as shown in the right-hand side panel of fig. \ref{p_mu_b_A_grp_A05}; a low aspiration level ($A = 0.1$) even led to the enhancement of cooperation. 

\begin{figure}[tb]
\centering
\vspace{0mm}
\includegraphics[width = 85mm, trim= 0 10 0 0]{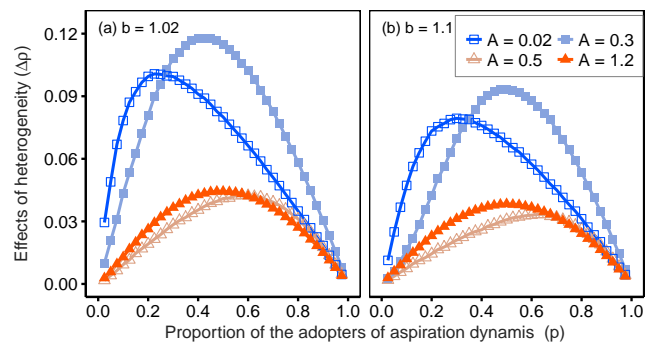}
\caption{\small The effect of heterogeneity as a function of the proportion of adopters of aspiration dynamics ($p$). A positive nonlinear effect was observed, particularly with a low aspiration level. Fixed parameters: $L = 100, \beta = 10, \mu = 0.1$.}
\label{p_A_b_dif}
\end{figure}

\begin{figure}[tb]
\centering
\vspace{0mm}
\includegraphics[width = 85mm, trim= 0 10 0 0]{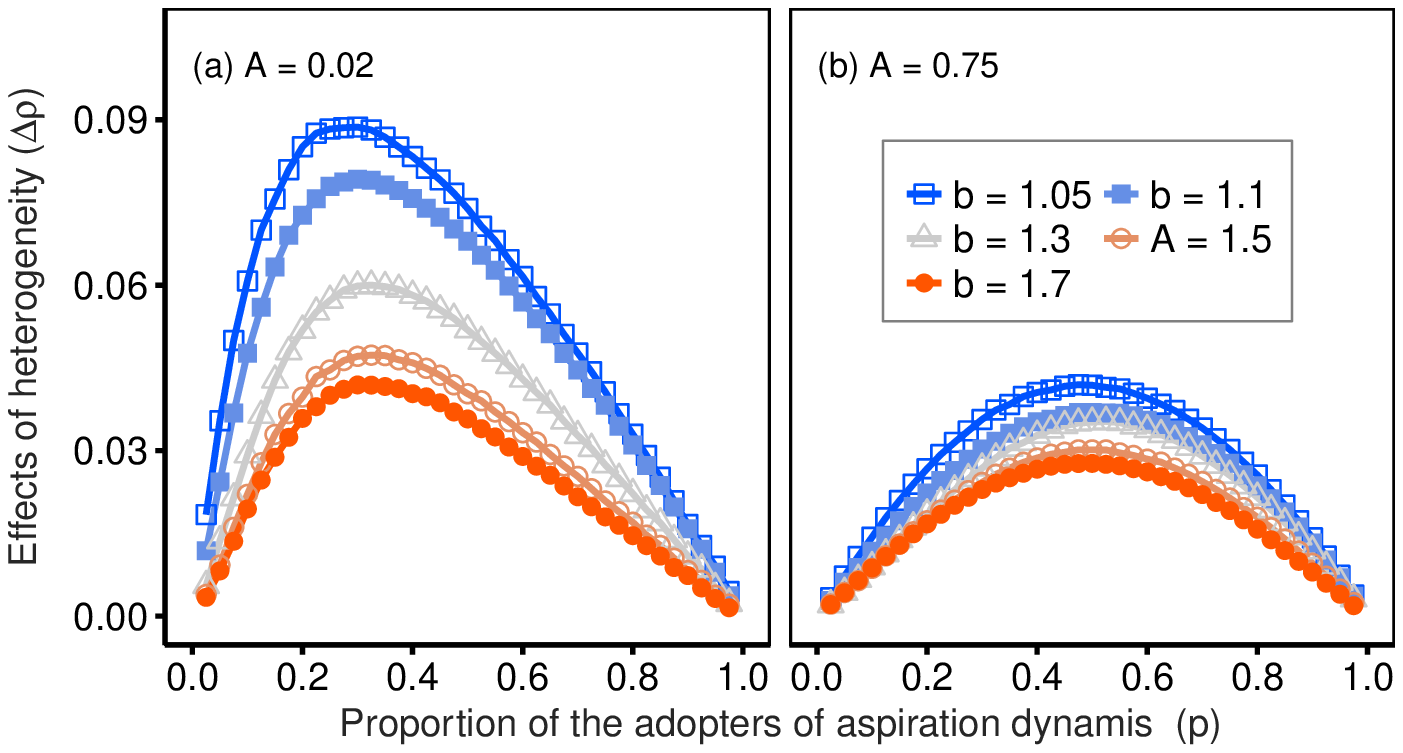}
\caption{\small The effect of heterogeneity as a function of the proportion of adopters of aspiration dynamics ($p$). The qualitative patterns observed in the fig. \ref{p_A_b_dif} were replicated with various values of $b$. Fixed parameters: $L = 100, \beta = 10, \mu = 0.1$.}
\label{p_b_A_dif}
\end{figure}

Last, using other values of $A$, we examined how cooperation depends on the aspiration level. Figure \ref{p_A_b_dif} shows $\Delta \rho$ as a function of $p$ for different values of $A$ ($\mu$ is set at 0.1). The figure shows that the introduction of aspiration dynamics had a positive nonlinear effect, which further depended on the aspiration level, and cooperation was greatly enhanced with $A = 0.02$ and $A = 0.3$. In particular, when a large majority of the population comprised imitators (small $p$), lower values of $A$ led to the large enhancement of cooperation, which is in line with our argument above. Figure \ref{p_b_A_dif} shows that this pattern did not depend on the values of $b$. The larger $\Delta \rho$ was observed with small aspiration level ($A = 0.02$, panel (a)) than with large aspiration level ($A = 0.75$, panel (b)).

\section{Discussion}
We conducted Monte Carlo simulations to examine the role of innovation in social dilemma by paying attention to the mutation rate and heterogeneous strategy updating rules. Our simulation results demonstrate that in combination with heterogeneous updating rules, mutation causes a deterioration in cooperation if the mutation rate is not overly large. This detrimental effect is prominent if a majority of the population comprises imitators rather than adopters of aspiration dynamics. Consequently, heterogeneous populations achieve lower cooperation levels than the weighted average of homogeneous populations. However, frequent mutation and heterogeneous dynamics sometimes foster cooperation relative to the average results of homogeneous systems. When cooperation among imitators is destroyed by a high mutation rate, the introduction of aspiration dynamics has positive nonlinear effects on cooperation. This effect is prominent when the introduced agents have low aspiration levels.

Although our study contributes to an understanding of the role of innovation by examining the effects of updating rules and mutation rates, there is room for further investigation. Considering other updating rules---such as Glauber dynamics and conformity---is an obvious line of further research. Even among payoff-based copying rules, various types of updating rules may be considered \cite{Szolnoki2018c}. Imitation process can be enriched by including other factors, such as emotion \cite{Szolnoki2011c, Szolnoki2013}. Our understanding concerning the emergence of cooperation will be enriched by investigation into the combination of these rules and their vulnerability to mutation. Another possible extension is the coevolution of updating rules. Here, the proportion of each rule-updating strategy is not fixed; it can coevolve with strategy on the basis of the fitness of each rule. The coevolutionary mechanism also leads to the expansion and shrinking of the parameter region where cooperation is sustained \cite{Moyano2009, Cardillo2010, Xu2017a, Danku2018, Szolnoki2018a}. Understanding the manner in which mutation affects this coevolutionary process is an interesting topic for future research---one that may help us understand why some updating rules are adopted by actual agents, including humans. Furthermore, the application of the framework of this paper and related literature will ultimately contribute to an understanding of other types of moral behaviour beyond cooperation \cite{Sinatra2009, Szolnoki2012a, Takesue2017a, Capraro2018}.


\begin{thebibliography}{10}
\expandafter\ifx\csname url\endcsname\relax\def\url#1{\texttt{#1}}\fi

\bibitem{Hardin1968}
\Name{Hardin G.} \REVIEW{Science}{162}{1968}{1243}.

\bibitem{Axelrod1984}
\Name{Axelrod R.} \Book{{The evolution of cooperation}} (Basic Books, New
York) 1984.

\bibitem{Nowak2006}
\Name{Nowak M.~A.} \REVIEW{Science}{314}{2006}{1560}.

\bibitem{Szabo2007a}
\Name{Szab{\'{o}} G. \and F{\'{a}}th G.} \REVIEW{Phys. Rep.}{446}{2007}{97}.

\bibitem{Roca2009}
\Name{Roca C.~P., Cuesta J.~A. \and S{\'{a}}nchez A.} \REVIEW{Phys. Life
Rev.}{6}{2009}{208}.

\bibitem{Perc2017}
\Name{Perc M., Jordan J.~J., Rand D.~G., Wang Z., Boccaletti S. \and Szolnoki
A.} \REVIEW{Phys. Rep.}{687}{2017}{1}.

\bibitem{Sanchez2018}
\Name{S{\'{a}}nchez A.} \REVIEW{J. Stat. Mech}{}{2018}{024001}.

\bibitem{Nowak1992}
\Name{Nowak M.~A. \and May R.~M.} \REVIEW{Nature}{359}{1992}{826}.

\bibitem{Santos2005}
\Name{Santos F.~C. \and Pacheco J.~M.} \REVIEW{Phys. Rev.
Lett.}{95}{2005}{098104}.

\bibitem{Szolnoki2008}
\Name{Szolnoki A., Perc M. \and Danku Z.} \REVIEW{Physica A}{387}{2008}{2075}.

\bibitem{Roca2009a}
\Name{Roca C.~P., Cuesta J.~A. \and S{\'{a}}nchez A.} \REVIEW{Phys. Rev.
E}{80}{2009}{046106}.

\bibitem{Zimmermann2004}
\Name{Zimmermann M.~G., Egu{\'{i}}luz V.~M. \and {San Miguel} M.} \REVIEW{Phys.
Rev. E}{69}{2004}{065102}.

\bibitem{Fu2007}
\Name{Fu F., Chen X., Liu L. \and Wang L.} \REVIEW{Physica A}{383}{2007}{651}.

\bibitem{Szolnoki2008b}
\Name{Szolnoki A., Perc M. \and Danku Z.} \REVIEW{EPL}{84}{2008}{50007}.

\bibitem{Perc2010}
\Name{Perc M. \and Szolnoki A.} \REVIEW{Biosystems}{99}{2010}{109}.

\bibitem{Wardil2014}
\Name{Wardil L. \and Hauert C.} \REVIEW{Sci. Rep.}{4}{2015}{5725}.

\bibitem{Helbing2010c}
\Name{Helbing D., Szolnoki A., Perc M. \and Szab{\'{o}} G.} \REVIEW{PLoS
  Comput. Biol.}{6}{2010}{e1000758}.

\bibitem{Szolnoki2013d}
\Name{Szolnoki A. \and Perc M.} \REVIEW{J. Theor. Biol.}{325}{2013}{34}.

\bibitem{Chen2014}
\Name{Chen X., Szolnoki A. \and Perc M.} \REVIEW{New J.
  Phys.}{16}{2014}{083016}.

\bibitem{Chen2015c}
\Name{Chen X., Sasaki T., Brannstrom A. \and Dieckmann U.} \REVIEW{J. R. Soc.
Interface}{12}{2015}{20140935}.

\bibitem{Chen2015}
\Name{Chen X., Szolnoki A. \and Perc M.} \REVIEW{Phys. Rev.
E}{92}{2015}{012819}.

\bibitem{Perc2015}
\Name{Perc M. \and Szolnoki A.} \REVIEW{Sci. Rep.}{5}{2015}{11027}.

\bibitem{Yang2015b}
\Name{Yang H.-X., Wu Z.-X., Rong Z. \and Lai Y.-C.} \REVIEW{Phys. Rev.
E}{91}{2015}{022121}.

\bibitem{Ohdaira2016}
\Name{Ohdaira T.} \REVIEW{Sci. Rep.}{6}{2016}{25413}.

\bibitem{Ohdaira2017}
\Name{Ohdaira T.} \REVIEW{Sci. Rep.}{}{2017}{12448}.

\bibitem{Takesue2018}
\Name{Takesue H.} \REVIEW{EPL}{121}{2018}{48005}.

\bibitem{Yang2010}
\Name{Yang H.-X., Wu Z.-X. \and Wang B.-H.} \REVIEW{Phys. Rev.
E}{81}{2010}{065101}.

\bibitem{Lin2011}
\Name{Lin Y.-T., Yang H.-X., Wu Z.-X. \and Wang B.-H.} \REVIEW{Physica A}{390}{2011}{77}.

\bibitem{Amaral2015}
\Name{Amaral M.~A., Wardil L. \and da~Silva J. K.~L.} \REVIEW{J. Phys. A Math.
Theor.}{48}{2015}{445002}.

\bibitem{Amaral2016a}
\Name{Amaral M.~A., Wardil L., Perc M. \and da~Silva J. K.~L.} \REVIEW{Phys.
Rev. E}{93}{2016}{042304}.

\bibitem{Li2015a}
\Name{Li K., Cong R., Wu T. \and Wang L.} \REVIEW{Phys. Rev.
E}{91}{2015}{042810}.

\bibitem{Li2016d}
\Name{Li K., Cong R. \and Wang L.} \REVIEW{EPL}{114}{2016}{58001}.

\bibitem{Szolnoki2017}
\Name{Szolnoki A. \and Chen X.} \REVIEW{Phys. Rev. E}{95}{2017}{052316}.

\bibitem{Huang2017b}
\Name{Huang C. \and Dai Q.} \REVIEW{EPL}{118}{2017}{28002}.

\bibitem{Huang2018f}
\Name{Huang C., Dai Q. \and Li H.} \REVIEW{EPL }{124}{2018}{18001}.

\bibitem{Szolnoki2018b}
\Name{Szolnoki A. \and Chen X.} \REVIEW{Phys. Rev. E}{98}{2018}{022309}.

\bibitem{Szolnoki2017b}
\Name{Szolnoki A. \and Chen X.} \REVIEW{EPL}{120}{2017}{58001}.

\bibitem{Chen2018c}
\Name{Chen X. \and Szolnoki A.} \REVIEW{PLoS Comput.
Biol.}{14}{2018}{e1006347}.

\bibitem{Szabo1998}
\Name{Szab{\'{o}} G. \and T{\H o}ke C.} \REVIEW{Phys. Rev. E}{58}{1998}{69}.

\bibitem{Rand2013}
\Name{Rand D.~G., Tarnita C.~E., Ohtsuki H. \and Nowak M.~A.} \REVIEW{Proc.
Natl. Acad. Sci.}{110}{2013}{2581}.

\bibitem{Chen2008}
\Name{Chen X. \and Wang L.} \REVIEW{Phys. Rev. E}{77}{2008}{017103}.

\bibitem{Zhang2011a}
\Name{Zhang J., Fang Y.~P., Du W.~B. \and Cao X.~B.} \REVIEW{Physica A}{390}{2011}{2258}.

\bibitem{Yang2012a}
\Name{Yang H.-X., Rong Z., Lu P.-M. \and Zeng Y.-Z.} \REVIEW{Physica A}{391}{2012}{4043}.

\bibitem{Du2014}
\Name{Du J., Wu B., Altrock P.~M. \and Wang L.} \REVIEW{J. R. Soc.
Interface}{11}{2014}{20140077}.

\bibitem{Amaral2016}
\Name{Amaral M.~A., Wardil L., Perc M. \and da~Silva J. K.~L.} \REVIEW{Phys.
Rev. E}{94}{2016}{032317}.

\bibitem{Roca2009b}
\Name{Roca C.~P., Cuesta J.~A. \and S{\'{a}}nchez A.} \REVIEW{Eur. Phys. J.
B}{71}{2009}{587}.

\bibitem{Li2017b}
\Name{Li Y.} \REVIEW{Phys. Rev. E}{95}{2017}{022303}.

\bibitem{Amaral2017}
\Name{Amaral M.~A., Perc M., Wardil L., Szolnoki A., {da Silva J{\'{u}}nior}
E.~J. \and da~Silva J. K.~L.} \REVIEW{Phys. Rev. E}{95}{2017}{032307}.

\bibitem{Amaral2018}
\Name{Amaral M.~A. \and Javarone M.~A.} \REVIEW{Phys. Rev.
E}{97}{2018}{042305}.

\bibitem{Danku2018}
\Name{Danku Z., Wang Z. \and Szolnoki A.} \REVIEW{EPL}{121}{2018}{18002}.

\bibitem{Pena2009}
\Name{Pe{\~{n}}a J., Volken H., Pestelacci E. \and Tomassini M.} \REVIEW{Phys.
Rev. E}{80}{2009}{016110}.

\bibitem{Cui2013}
\Name{Cui P.-B. \and Wu Z.-X.} \REVIEW{Physica A}{392}{2013}{1500}.

\bibitem{Szolnoki2015}
\Name{Szolnoki A. \and Perc M.} \REVIEW{J. R. Soc.
Interface}{12}{2015}{20141299}.

\bibitem{Javarone2016a}
\Name{Javarone M.~A., Antonioni A. \and Caravelli F.} \REVIEW{EPL}{114}{2016}{38001}.

\bibitem{Shu2019}
\Name{Shu F., Liu Y., Liu X. \and Zhou X.} \REVIEW{Appl. Math.
Comput.}{346}{2019}{480}.

\bibitem{Szolnoki2018c}
\Name{Szolnoki A. \and Danku Z.} \REVIEW{Physica A}{511}{2018}{371}.

\bibitem{Liu2016b}
\Name{Liu X., He M., Kang Y. \and Pan Q.} \REVIEW{Phys. Rev.
E}{94}{2016}{012124}.

\bibitem{Xu2017a}
\Name{Xu K., Li K., Cong R. \and Wang L.} \REVIEW{EPL}{117}{2017}{48002}.

\bibitem{Macy2015}
\Name{Macy M. \and Tsvetkova M.} \REVIEW{Sociol. Methods Res.}{44}{2015}{306}.

\bibitem{Helbing2009}
\Name{Helbing D. \and Yu W.} \REVIEW{Proc. Natl. Acad. Sci.}{106}{2009}{3680}.

\bibitem{Ichinose2018a}
\Name{Ichinose G., Satotani Y. \and Sayama H.} \REVIEW{New J.
Phys.}{20}{2018}{053049}.

\bibitem{Adami2016}
\Name{Adami C., Schossau J. \and Hintze A.} \REVIEW{Phys. Life
Rev.}{19}{2016}{1}.

\bibitem{Du2015}
\Name{Du J., Wu B. \and Wang L.} \REVIEW{Sci. Rep.}{5}{2015}{8014}.

\bibitem{Huang2017}
\Name{Huang C., Dai Q., Cheng H. \and Li H.} \REVIEW{EPL}{120}{2017}{18001}.

\bibitem{Takesue2019}
\Name{Takesue H.} \REVIEW{Physica A}{513}{2019}{399}.

\bibitem{Szolnoki2011c}
\Name{Szolnoki A., Xie N.-g., Wang C. \and Perc M.} \REVIEW{EPL}{96}{2011}{38002}.

\bibitem{Szolnoki2013}
\Name{Szolnoki A., Xie N.-G., Ye Y. \and Perc M.} \REVIEW{Phys. Rev.
  E}{87}{2013}{042805}.

\bibitem{Moyano2009}
\Name{Moyano L.~G. \and S{\'{a}}nchez A.} \REVIEW{J. Theor.
Biol.}{259}{2009}{84}.

\bibitem{Cardillo2010}
\Name{Cardillo A., G{\'{o}}mez-Garde{\~{n}}es J., Vilone D. \and S{\'{a}}nchez
A.} \REVIEW{New J. Phys.}{12}{2010}{103034}.

\bibitem{Szolnoki2018a}
\Name{Szolnoki A. \and Chen X.} \REVIEW{New J. Phys.}{20}{2018}{093008}.

\bibitem{Sinatra2009}
\Name{Sinatra R., Iranzo J., G{\'{o}}mez-Garde{\~{n}}es J., Flor{\'{i}}a L.~M.,
Latora V. \and Moreno Y.} \REVIEW{J. Stat. Mech.}{2009}{2009}{P09012}.

\bibitem{Szolnoki2012a}
\Name{Szolnoki A., Perc M. \and Szab{\'{o}} G.} \REVIEW{EPL}{100}{2012}{28005}.

\bibitem{Takesue2017a}
\Name{Takesue H., Ozawa A. \and Morikawa S.} \REVIEW{EPL}{118}{2017}{48002}.

\bibitem{Capraro2018}
\Name{Capraro V. \and Perc M.} \REVIEW{Front. Phys.}{6}{2018}{107}.

\end{thebibliography}

\end{document}